\documentclass[aps,prl,twocolumn]{revtex4}
\usepackage{array}
\usepackage{amsmath}
\usepackage{graphicx}
\usepackage{amssymb}
\usepackage{latexsym}

\newcommand{\bra}[1]{\ensuremath{\left\langle {#1} \right|}}
\newcommand{\ket}[1]{\ensuremath{\left| {#1} \right\rangle}}

\renewcommand{\vec}[1]{\mathbf{#1}}

\begin{document}

\title{Uniform Peak Optical Conductivity in Single-Walled Carbon Nanotubes}

\author{Jesse M. Kinder}
\email{jk959@cornell.edu}
\affiliation{Department of Chemistry and Chemical Biology, Cornell University,
Ithaca, New York 14853}

\author{Jiwoong Park}
\affiliation{Department of Chemistry and Chemical Biology, Cornell University,
Ithaca, New York 14853}
\affiliation{Kavli Institute at Cornell for Nanoscale Science, Ithaca, New York
14853,USA.}

\author{Garnet Kin-Lic Chan}
\affiliation{Department of Chemistry and Chemical Biology, Cornell University,
Ithaca, New York 14853}

\begin{abstract}

	Recent measurements in single-walled carbon nanotubes show that, on
	resonance, all nanotubes display the same peak optical conductivity of
	approximately 8 $e^2/h$, independent of radius or chirality [Joh \emph{et
	al.}, \emph{Nature Nanotechnology} \textbf{6}, 51 (2011)]. We show that this
	uniform peak conductivity is a consequence of the relativistic band
	structure and strength of the Coulomb interaction in carbon nanotubes.  We
	further construct a minimalist model of exciton dynamics that describes the
	general phenomenology and provides an accurate prediction of the numerical
	value of the peak optical conductivity. The work illustrates the need for
	careful treatment of relaxation mechanisms in modeling the optoelectronic
	properties of carbon nanotubes.
	
\end{abstract}

\date{\today}

\maketitle

\begin{figure}
\centerline{\includegraphics[width=1.0\columnwidth]{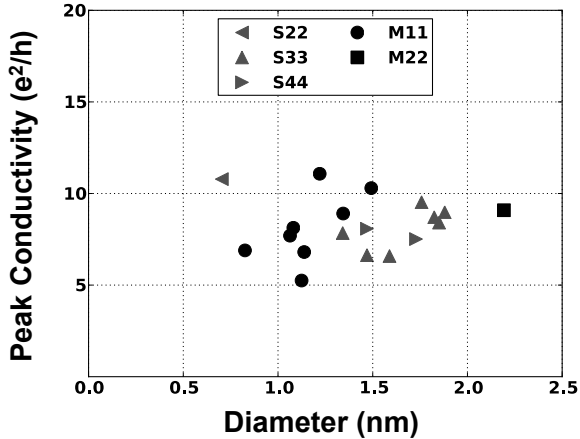}}

\caption{Optical conductivity on resonance as a function of nanotube diameter,
as measured by Joh, \emph{et al.}~\cite{joh2010unp}. The mean value is $8 \pm 1.5
\, e^2 / h$, independent of the diameter. }
\label{fig:data}

\end{figure}

Using a new on-chip Rayleigh scattering technique, Joh \emph{et
al.}~measured the peak optical conductivity for a variety of transitions in
semiconducting and metallic nanotubes \cite{joh2010ocr,joh2010unp}. The data
reveal a surprising phenomenon: on resonance, the optical conductivity is
independent of the nanotube radius and narrowly distributed around $\sigma_*
= 8 \, e^2 / h$. On resonance, \emph{all} nanotubes respond like classical
conducting hollow cylinders with the \emph{same} conductivity. The sample
included the second, third, and fourth exciton transitions in semiconducting
nanotubes ($S_{22}$, $S_{33}$ and $S_{44}$) and the first and second exciton
transitions in metallic nanotubes ($M_{11}$ and $M_{22}$). The data are shown in
Fig.~\ref{fig:data}. 

In this Letter we identify the origin of this uniform peak conductivity. We analyze
the optical conductivity within linear response theory for the effective Dirac
model of a carbon nanotube. Our main result is that the peak conductivity will
be independent of the nanotube radius $R$ whenever quasiparticle energies are
proportional to $1/R$ and quasiparticle lifetimes are proportional to $R$. The
Coulomb interaction satisfies both requirements, and a simple exciton model with
interband Coulomb scattering fits the data quite well. We also consider the
effects of phonon and impurity scattering, which may account for some of the
spread in the data. The analysis suggests that a uniform peak conductivity
should be observable over a wide range of experimental conditions and
illustrates the importance of relaxation mechanisms in numerical models of
nanotube properties.

Our starting point is the low-energy approximation to the tight-binding
Hamiltonian for a carbon nanotube, which is a massless Dirac equation
\cite{kane1997ssa}:
\begin{equation}
	\mathcal{H} = \hbar v_F \, \vec{q} \cdot \boldsymbol{\sigma}.
	\label{eq:dirac}
\end{equation}
$v_F$ is the Fermi velocity and $\vec{q}$ describes small displacements from the
corners of the Brillouin zone of graphene. $R$ is the only relevant length
scale, and it defines a natural energy scale:
\begin{equation}
	E_0 = \hbar v_F/R .
\end{equation}
This defines an effective mass as well: $E_0 = m^{*}v_F{}^2$. The eigenvalues of
$\mathcal{H}$ are $\pm E_0 \sqrt{ (kR)^2 + \Delta^2}$, where $k$ is the wave
vector along the nanotube axis and $\Delta$ is the gap parameter.  In metallic
nanotubes, $\Delta = N$, and in semiconducting nanotubes, $\Delta = N \pm 1/3$.
The band index $N$ is an integer.

This free particle model can be extended to include the Coulomb interaction
\cite{ando1997ecn}. In a single-band model, exciton wave functions are of the
form
\begin{equation}
	\ket{n} = \sum_{k} A_{n,k}
		\cdot c_{k}^{\dag} \, v_{k}^{\,} \ket{\Omega}.
	\label{eq:waveFunction}
\end{equation}
$c^\dag$ creates a conduction electron, $v^\dag$ creates a valence electron, and
$\ket{\Omega}$ is the ground state of filled valence orbitals.  The coefficients
$A_{n,k}$ and exciton energies $E_n$ are obtained as eigenvectors and
eigenvalues of a Bethe Salpeter equation. This single-band model is sufficient
to describe direct excitons ($S_{ii}$ and $M_{ii}$) in carbon nanotubes
\cite{dresselhaus2007epc}, and the scaling of exciton size and binding
energy obtained from this model and \emph{ab initio} calculations agree to
leading order in $R$ \cite{perebeinos2004sec, capaz2006dac}.

At the level of linear response theory, the conductivity is given by the Kubo
formula:
\begin{equation}
	\sigma(\omega) = 
	\dfrac{\hbar e^2}{i 2 \pi R L} \sum_{n}
	\dfrac{ | \bra{ n } \hat{v} \ket{\Omega} |^2 } { E_n }
	\dfrac{ 2 (\hbar \omega + i\hbar \Gamma_{n}) }
	{ E_n{}^2 - (\hbar \omega + i\hbar \Gamma_{n})^2 } .
	\label{eq:kuboI}
\end{equation}
$\hat{v}$ is the velocity operator projected along the nanotube axis.  $\Gamma_n
= 1/\tau_n$ where $\tau_n$ is the lifetime of the excited state.  If $\ket{n}$
were eigenstates of the Hamiltonian in the absence of an applied field,
$\Gamma_{n}$ could be replaced by an infinitesimal to enforce causality.  In
practice, the exact eigenstates are never known. A broadening parameter $1/\tau$
is often introduced to account for relaxation. However, $\tau$ is not
arbitrary: different scattering mechanisms lead to qualitatively different
lifetimes, and the broadening determines the optical conductivity on resonance.

Eq.~(\ref{eq:kuboI}) gives the conductivity of a single band. The total
conductivity is multiplied a factor of 2 for spin and a factor of 2 for the $K$
and $K'$ points. Although the conductivity is a tensor, scattering in nanotubes
is dominated by light polarized along the nanotube axis \cite{islam2004dmp}.
Only the surface current along the axis due to an electric field applied along
the axis is considered here: $\sigma(\omega) \equiv \sigma_{zz}(\omega)$.

The Kubo formula may be simplified by scaling all energies by $E_0$ and
rewriting the expression in terms of the dimensionless parameters
\begin{align}
	w &= \hbar \omega / E_0 , \notag \\
	x_n &= E_n / E_0 , \label{eq:dimensionless} \\
	y_n &= \hbar \Gamma_n / E_0. \notag
\end{align}
Eq.~(\ref{eq:kuboI}) may be simplified further by introducing a dimensionless oscillator strength
\cite{ando1997ecn}:
\begin{equation}
	f_n = 2m^* \dfrac{ | \bra{ n } \hat{v} \ket{\Omega} |^2 } { E_n } .
	\label{eq:fOscillator}
\end{equation}

The oscillator strengths for the solutions of the Dirac Hamiltonian satisfy a
sum rule \footnote{This follows from Eq.~(\ref{eq:fOscillator}) when the sum over
$k$ is approximated by an integral.}: $\sum_k f_k = L/\pi R$.  For an exciton,
$f_n = \phi_n \cdot L/\pi R$, where the fractional oscillator strength $\phi_n$
is independent of $R$.

After these substitutions, Eq.~(\ref{eq:kuboI}) becomes
\begin{equation}
	\sigma(\omega) = 
	\dfrac{e^2}{h} \dfrac{1}{i \pi} \sum_{n} \phi_n
	\cdot
		\dfrac{ w + iy_{n} }
		{ x_n{}^2 - (w + iy_{n})^2 } .
	\label{eq:kuboII}
\end{equation}

This expression implies $\sigma(\omega) = G(w; \boldsymbol{\lambda}) \cdot e^2/h
$, where $G$ is a dimensionless function and $\boldsymbol{\lambda}$ is a set of
dimensionless parameters derived from $\{x_n\}$ and $\{y_n\}$. On resonance, $w$
is a function of the other parameters so that $\sigma_* =
G_*(\boldsymbol{\lambda}) \cdot e^2/h$.  If $x_n$ and $y_n$ are independent of
the nanotube radius, then so is the peak conductivity.
Eq.~(\ref{eq:dimensionless}) implies that $x_n$ and $y_n$ are independent of the
nanotube radius if $E_n$ and $\Gamma_n$ are proportional to $1/R$. This proves
our central result:

\emph{If quasiparticle energies are inversely proportional to the nanotube
radius and quasiparticle lifetimes are proportional to the nanotube radius,
then the conductivity on resonance is independent of the nanotube radius.}

The arguments of this section can be extended to multiple bands and indirect
excitations. The result also holds for unbound electron-hole pairs \footnote{Although the
fractional oscillator strength is no longer proportional to $L/R$, transforming
the sum in Eq.~(\ref{eq:kuboI}) into an integral over $k$ leads to the same
dependence on $w$ and $\boldsymbol{\lambda}$.}.

The first requirement is satisfied as long as Eq.~(\ref{eq:dirac}) is valid. If
the Dirac equation is scaled by $E_0$, then $k$ only enters in the dimensionless
combination $kR$. The scaled Coulomb interaction only depends on an effective
fine structure constant, $\alpha = e^2 / 2 \pi \kappa \hbar v_F$ where $\kappa$
describes static screening from the environment and other bands of the nanotube.
As a result, the scaled Hamiltonian only depends on the dimensionless variables
$\xi=kR$ and $\alpha$. The spectrum and eigenvectors are universal functions of
these variables, and the energy eigenvalues for both free particles and excitons
are proportional to $\hbar v_F / R$. This inverse relation between the energy
and the radius is a consequence of the relativistic band structure of a carbon
nanotube, and the conditions that lead to a uniform peak conductivity in carbon
nanotubes are not satisfied in a general quantum wire.

The second requirement is not satisfied in general. Quasiparticle lifetimes
arise from interactions not included in the Dirac equation, and each interaction
must be analyzed separately to determine whether the lifetime is proportional to
$R$. The lifetime of a state $\ket{n}$ due to a potential $V$ can be estimated
from Fermi's golden rule:
\begin{align}
	\Gamma_n &= (2\pi/\hbar) \sum_{m} |V_{m,n}|^2 \cdot \delta(E_m - E_n) .
	\label{eq:fgr}
\end{align}
All of the dimensional quantities can be collected into a base scattering rate
$\Gamma_0$. The remaining sum defines a dimensionless function $\gamma$ whose
analytic form is irrelevant to the main result. The scattering rate is then
$\gamma \cdot \Gamma_0$. If $\Gamma_0 \propto 1/R$ and $\gamma$ is independent
of $R$, the quasiparticle lifetime will satisfy the requirements for a
radius-independent peak conductivity.

The Coulomb interaction largely determines the photophysics of carbon nanotubes.
The absorption of a photon produces a particle-hole pair, and these charged
particles interact strongly through their mutual Coulomb attraction. The
interaction between particles and holes in the same band leads to strong exciton
binding. Interband scattering gives the exciton a finite lifetime
\cite{kane2003rps}. The base rate is
\begin{equation}
	\hbar \Gamma_0 = \alpha^2 E_0,
	\label{eq:dissociation}
\end{equation}
and $\gamma$ is independent of $R$. Thus, interband Coulomb scattering
leads to a peak conductivity that is independent of the radius.

In fact, an exciton model with dissociation due to interband Coulomb scattering
accounts for all of the qualitative features of the data: the peak conductivity
is independent of the radius, does not depend strongly on the transition
responsible for the resonance, and is approximately equal in semiconducting
and metallic nanotubes.

Dissociation rates show little variation between bands. The lifetime of an
exciton is approximately equal to that of a particle-hole pair at the band edge
\cite{kane2003rps}. For the $S_{33}$, $S_{44}$, $M_{11}$, and $M_{22}$ excitons,
$\gamma$ falls between 0.7 and 1.2. The estimated scattering rate of the
$S_{22}$ exciton is significantly larger because of the small overlap of wave
functions in the first and second bands \footnote{The single $S_{22}$ transition
in Fig.~\ref{fig:data} has a peak conductivity of $10.8 \, e^2/h$. This is
smaller than predicted by the Dirac model, but corrections will be significant
because the nanotube radius is so small (0.35 nm).}.

The bands in metallic nanotubes are twofold degenerate, which suggests the peak
conductivity of metallic nanotubes could be more than twice that of
semiconducting nanotubes. However, screening in metallic nanotubes reduces the
exciton binding energy and oscillator strength. Trigonal warping lifts the
degeneracy and further reduces the conductivity in chiral nanotubes. These
factors lead to similar peak conductivities for semiconducting and metallic
nanotubes.

Dissociation due to interband Coulomb scattering also accounts for the magnitude
of the peak conductivity. If all the oscillator strength of a band is localized
in a single transition and $\gamma \approx 1$, the peak conductivity is
\begin{equation}
	\sigma_* = \dfrac{e^2}{h} \cdot \dfrac{2\kappa^2}{\pi \alpha_0{}^2},
\end{equation}
where $\alpha_0 = e^2/2\pi\hbar v_F \approx 0.42$, and $\kappa$ is the
dielectric constant of the environment. The experimental measurements shown in
Fig.~\ref{fig:data} were taken on a quartz substrate in glycerol ($n=1.46$).
Setting $\kappa = n^2$ gives a maximum conductivity of $\sigma_* \approx 16 \,
e^2/h$. A peak conductivity of $8 \, e^2/h$ follows if the exciton transition
accounts for half the total oscillator strength, a fraction consistent with
\emph{ab initio} calculations \cite{perebeinos2004sec}.

Other interactions also contribute to the exciton lifetime and may account for
some of the spread in the data of Fig.~\ref{fig:data}. Intrinsic sources of
scattering include phonons and lattice defects.  External perturbations such as
the substrate, applied fields, or atoms adsorbed on the surface of the nanotube
also affect the conductivity. Here we consider phonon and impurity scattering
to illustrate how different scattering mechanisms lead to qualitatively
different quasiparticle lifetimes.

Electron-phonon interactions are a significant source of scattering in
carbon nanotubes \cite{park2004eps, zhou2005bsp, dresselhaus2005rsc,
perebeinos2005eep, miyauchi2006iep}. Lattice deformations introduce an effective
potential to Eq.~(\ref{eq:dirac}). A general deformation has two effects
\cite{kane1997ssa, suzuura2002pae}. First, variations in the lattice charge
density can produce a scalar deformation potential. Second, bending and
stretching of bonds can introduce an effective gauge potential. The orientation
of the bonds with respect to the axis of a nanotube depends on its chirality, 
and phonon scattering rates depend on the chiral angle $\theta_c$.

There are four acoustic modes in carbon nanotubes. The longitudinal acoustic
(LA) modes compress and expand the lattice along the nanotube axis. The
transverse acoustic (TA) or twist modes rotate the lattice about the nanotube
axis. Both of these modes have a linear dispersion relation: $\omega_q = c q$
where $c$ is the sound velocity. These modes do not carry any angular momentum
and lead to small-momentum scattering within a band. The two flexure modes
bend the nanotube in the plane of its axis. These modes have a quadratic
dispersion relation: $\omega_q = c q^2 R$ where $c$ has units of velocity. These
modes carry one quantum of angular momentum and mediate interband transitions.

A radial breathing mode (RBM) is a periodic variation in the radius of a nanotube
along its axis. Its frequency is inversely proportional to the nanotube radius.
At small wave vector, $\omega_q \approx c/R$, where $c$ has units of velocity.
In nanotube with a 1 nm diameter, the energy of the lowest RBM is about 27
meV \cite{saito1999ppc}.

The scattering rates of the RBM, LA, TA, and flexure modes all have the same
form.  At zero temperature,
\begin{equation}
	\hbar \Gamma_{0} =
	\dfrac{ g{}^2 }{ \rho_0 c v_F R^2 }.
\end{equation}
In the high-temperature limit,
\begin{equation}
	\hbar \Gamma_{0} =
	\dfrac{ g{}^2 k_B T}{ \hbar \rho_0 c^2 v_F R } .
\end{equation}
$g$ characterizes the electron-phonon interaction strength, $T$ is the
temperature, and $\rho_0$ is the mass density of the graphene lattice. Although
a single parameter $g$ appears above, the deformation potential is an order of
magnitude larger than the gauge potential \cite{suzuura2002pae}. As a result,
the LA and RB modes, which contribute to the deformation potential, generally
have larger scattering rates than the other modes.  $\gamma$ depends on $c/v_F$ and
$\theta_c$, but not $R$. The contribution of these modes to the peak
conductivity is independent of the radius in the high-temperature limit. In
contrast, optical modes do \emph{not} contribute to a uniform peak conductivity.

Analyzing variations in the peak conductivity with temperature would provide a
way to extract the phonon contribution. The contribution of the RBM should show
a crossover to radius-dependent scaling as the temperature is reduced, and the
contribution of the acoustic modes should scale linearly with temperature.

Impurities in the nanotube or its environment provide another source of
scattering. A short-range potential localized on the surface of a
nanotube can be approximated by a point-like impurity potential: $V(\vec{r})
\approx V_0 \, a^2 \, \delta(\vec{r})$, where $a \alt R$ is the range of the
potential. This might represent a topological defect in the lattice, a
substitution impurity at a lattice site, or an atom adsorbed on the surface of
the nanotube \footnote{A potential that affects the two sublattices
differently can be modeled by replacing $V_0$ with $V_0 \, \hat{\mathbf{n}}
\cdot \boldsymbol{\sigma}$. This does not affect the base scattering rate
$\Gamma_0$.}. Incoherent elastic scattering from identical impurities gives
\begin{equation}
	\hbar \Gamma_0 = \dfrac{ \rho a^4 V_0{}^2}{\hbar v_F R} ,
\end{equation}
where $\rho$ is the surface defect density.

A long-range potential that is nearly uniform around the circumference of a
nanotube but localized along its axis may be approximated by a one-dimensional
impurity potential: $V(\vec{r}) \approx a V_0 \, \delta(z)$, where $V_0$ is the
average potential around the circumference. This might represent a charge defect
in the substrate or a local gating potential. The resulting scattering rate is
proportional to the linear defect density and independent of $R$.

Elastic scattering from short-range impurities leads to a peak conductivity that
is independent of the radius; inelastic scattering and long-range impurity
scattering do not. The high mobilities observed in DC transport measurements
suggest that impurity scattering of any type is insignificant compared with
phonon scattering \cite{zhou2005bsp, park2004eps}.

The lifetime due to phonon and impurity scattering in metallic nanotubes
is twice as large as that in semiconducting nanotubes. Phonon and impurity
scattering also lead to stronger dependence on the band index than Coulomb
scattering. No general scaling arguments require the peak conductivity to be
independent of the band index. Although $\Delta \cdot E_0$ defines a natural
energy scale for a band, the peak conductivity depends on the gap
parameter even if $E_n$ and $\Gamma_n$ are proportional to $\Delta \cdot E_0$. A
simple model illustrates this. The peak conductivity can be calculated
analytically in a free particle model with a constant scattering rate.  If
$\Gamma_0 = c \Delta /R$, the peak conductivity is independent of the radius but
proportional to $1 / \Delta$. For phonon and impurity scattering, the lifetime
is independent of $\Delta$, and the peak conductivity is proportional to $1 /
\sqrt{\Delta}$. 

Phonon and impurity scattering may account for some of the spread in the data of
Fig.~\ref{fig:data}. Experimental error accounts for some variation. Acoustic
phonon scattering and short-range impurity scattering would introduce dependence
on the chiral angle; optical phonons, long-range impurities, and inelastic
scattering would introduce dependence on the radius. Other factors such as
doping or fluctuations in the local dielectric environment might also play a
role.

Trigonal warping may also account for some variation between nanotubes. The
Dirac Hamiltonian is a good approximation for a nanotube whose radius is much
larger than the lattice spacing. When the radius is small, trigonal warping
could lead to variations in the peak conductivity through its effect on
scattering rates, analogous to the family behavior of exciton binding energies
\cite{dresselhaus2007epc}.

To summarize, linear response theory predicts a peak conductivity that is
independent of the nanotube radius whenever quasiparticle energies are
inversely proportional to the nanotube radius and quasiparticle lifetimes are
proportional to the radius. The Coulomb interaction satisfies both
requirements and explains both the uniformity and mean value of the
conductivity data in Fig.~\ref{fig:data}. Phonon, impurity scattering, and
trigonal warping may account for small variations between nanotubes.

Our analysis suggests the peak conductivity will be \emph{uniform} over
a wide range of experimental conditions: it will be the same for all nanotubes
in a sample, independent of diameter or chirality. The peak conductivity is
not \emph{universal}, however, and may depend on factors such as the
dielectric environment, temperature, or doping. 

A uniform peak conductivity could be useful in optical devices that utilize
carbon nanotubes. Many properties of a nanotube depend strongly on its radius
or chirality. Applications designed to exploit these properties are faced
with the difficult task of separating nanotubes based on their geometry. In
applications that only depend on the resonant conductivity, nanotubes would be
interchangeable: on resonance, all nanotubes behave as classical wires with
the \emph{same} conductivity.

Our analysis also illustrates the importance of the broadening term in numerical
studies: a scattering rate inversely proportional to the nanotube radius is
essential to reproduce the uniform conductivity observed by Joh \emph{et al.} In
calculations of nanotube properties, it is common to introduce a
phenomenological parameter to account for scattering mechanisms not included in
the model. If the same parameter is used for different nanotubes, calculations
will yield incorrect scaling relations. To fit a data set, a separate parameter
could be adjusted for each nanotube. The experiments of Joh, \emph{et al}.~and
the analysis above suggest a different approach: for a given set of experimental
conditions, a single parameter can describe all nanotubes in a sample if the
lifetime is proportional to the nanotube radius: $\tau = \tau_0 \cdot (R/R_0)$.

J.M.K. would like to thank Daniel Joh and Lihong Hermann for many useful
discussions of the experiment.

This work was funded by
the Cornell Center for Materials Research,
the Center for Molecular Interfacing,
the NSF CAREER grant,
the Air Force Office of Scientific Research (NE and IO),
the Camille and Henry Dreyfus Foundation,
the David and Lucile Packard Foundation,
and the Alfred P. Sloan Foundation.

\end{document}